\documentclass{aa}  

\usepackage{graphicx}
\usepackage{txfonts}
\usepackage{hyperref}

\def\msun{{\rm ~M}_{\odot}}
\def\zsun{{\rm ~Z}_{\odot}}
\def\gpy{{\rm ~Gpc}^{-3} {\rm ~yr}^{-1}}

\begin{document}

   \title{The Effect of Pair-Instability Mass Loss on Black Hole Mergers}

   \author{K. Belczynski\inst{1}
          \and
          A. Heger\inst{2,3,4,5}
          \and
          W. Gladysz\inst{1}
          \and
          A. J. Ruiter\inst{6,7} 
          \and
          S. Woosley\inst{8}
          \and
          G. Wiktorowicz\inst{1}
          \and
          H.-Y. Chen\inst{9} 
          \and
          T. Bulik\inst{1}
          \and
          R. O'Shaughnessy\inst{10}
          \and
          D. E. Holz\inst{9}
          \and
          C. L. Fryer\inst{11}
          \and
          E. Berti\inst{12,13}          
   }

   \institute{Astronomical Observatory, Warsaw University, Ujazdowskie 4, 00-478 Warsaw, Poland\\
              \email{chrisbelczynski@gmail.com}
         \and
           Monash Centre for Astrophysics, School of Physics and Astronomy, 
           Monash University, Victoria 3800, Australia
         \and
           School of Physics \& Astronomy, University of Minnesota, 
           Minneapolis, MN 55455, USA
         \and
           Center for Nuclear Astrophysics, Department of Physics and Astronomy, 
           Shanghai Jiao-Tong University, Shanghai 200240, P. R. China
         \and
           Joint Institute for Nuclear Astrophysics, 1 Cyclotron Laboratory, 
           National Superconducting Cyclotron Laboratory, Michigan State University, 
           East Lansing, MI 48824-1321, USA
         \and 
           Research School of Astronomy and Astrophysics, The Australian National
           University, Canberra, ACT 2611, Australia
          \and
           ARC Centre of Excellence for All-sky Astrophysics (CAASTRO), Australia
         \and
           Department of Astronomy and Astrophysics, University of California,
           Santa Cruz, CA 95064, USA 
         \and
           Enrico Fermi Institute, Department of Physics, Department of
           Astronomy and Astrophysics, and Kavli Institute for Cosmological Physics, 
           University of Chicago, Chicago, IL 60637, USA
         \and
           Center for Computational Relativity and Gravitation, Rochester Institute of 
           Technology, Rochester, New York 14623, USA
         \and
           CCS-2, MSD409, Los Alamos National Laboratory, Los Alamos, NM 87545, USA
         \and
           Department of Physics and Astronomy, The University of Mississippi,
           University, MS 38677, USA
         \and 
           CENTRA, Departamento de F\'isica, Instituto Superior T\'ecnico,
           Universidade de Lisboa, Avenida Rovisco Pais 1, 1049 Lisboa, Portugal
   }

   \date{Received May 22, 2016; accepted ???}

 
  \abstract
   {Mergers of two stellar origin black holes are a prime source of gravitational 
    waves and are under intensive investigation.  One crucial ingredient in their 
    modeling has been neglected: pair-instability pulsation supernovae with 
    associated severe mass loss may suppress the formation of massive black holes, 
    decreasing black hole merger rates for the highest black hole masses.
   }
   {We demonstrate the effects of pair-instability pulsation supernovae on merger 
    rate and mass using populations of double black hole binaries formed through the 
    isolated binary classical evolution channel.    
   } 
   {The mass loss from pair-instability pulsation supernova is estimated based on
    existing hydrodynamical calculations. This mass loss is incorporated into 
    the {\tt StarTrack} population synthesis code. {\tt StarTrack} is used to 
    generate double black hole populations with and without pair-instability
    pulsation supernova mass loss.  
   }
   {The mass loss associated with pair-instability pulsation supernovae limits the 
    Population I/II stellar-origin black hole mass to $50\msun$, in tension with earlier 
    predictions that the maximum black hole mass could be as high as $100\msun$.
    In our model, neutron stars form with mass $1$--$2\msun$, then we encounter the 
    first mass gap at $2$--$5\msun$ with an absence of compact objects due to rapid supernova 
    explosions, followed by the formation of black holes with mass $5$--$50\msun$, with a 
    second mass gap at $50$--$135\msun$ created by pair-instability pulsation 
    supernovae and by pair-instability supernovae. Finally, black holes having masses above 
    $135\msun$ may {\em potentially} form to arbitrarily high mass limited only
    by the extent of the initial mass function and the strength of stellar winds.
    Suppression of double black hole merger rates by pair-instability pulsation 
    supernovae is negligible for our evolutionary channel. Our standard evolutionary 
    model with the inclusion of pair-instability pulsation supernovae and 
    pair-instability supernovae is fully consistent with the LIGO observations of 
    black hole mergers: GW150914, GW151226, and LVT151012. The LIGO results are inconsistent
    with high ($\gtrsim 400$ km s$^{-1}$) BH natal kicks. We predict the detection of 
    several, and up to as many as $\sim 60$, BH-BH mergers with a total mass of $10$--$150\msun$ 
    (most likely range: $20$--$80\msun$) in the forthcoming $\sim 60$ effective days of the LIGO 
    O2 observations, assuming the detectors reach the \emph{optimistic}\/ target O2 
    sensitivity. 
   }


   \maketitle

\section{Introduction}

In September 2015 the upgraded Laser Interferometer Gravitational-wave Observatory 
(LIGO) began observations with a sensitivity to the merger of two neutron stars to an 
average distance of $d_{\rm nsns} \approx 70$ Mpc. In this first upgraded science 
run (O1; \citet{LigoO1a,LigoO1b}) LIGO made two firm detections of 
black hole-black hole (BH-BH) mergers with component masses of $36\msun$ and $29\msun$ 
(GW150914), and $14\msun$ and $8\msun$ (GW151226), and also reported a candidate
BH-BH merger with masses of $23\msun$ and $13\msun$ (LVT151012). 
These discoveries verified earlier predictions that: \emph{(i)} the first detection 
would happen when LIGO sensitivity reaches $d_{\rm nsns}=50$--$100$ Mpc, that 
\emph{(ii)} BH-BH mergers would dominate the gravitational-wave signal, and that 
\emph{(iii)} the merging black holes would be substantially more massive than typical 
$10\msun$ Galactic BHs \citep{Belczynski2010a,Dominik2015,Belczynski2016a}.   

BH-BH mergers have been proposed as potential gravitational wave sources since the 1980s 
\citep{Bond1984,Thorne1987,Schutz1989} and have been studied since the 1990s 
\citep{Tutukov1993,Lipunov1997,Flanagan1998}. More recently, a number of groups 
have provided evolutionary models leading to potential BH-BH formation in 
a typical Galactic environment with high metallicity stars 
\citep{Brown2001,Nelemans2001,Belczynski2002,Voss2003,Postnov2006}.
Subsequently it was shown that BH-BH merger formation in the Galactic 
environment, with its high metallicity, leads to suppression of BH-BH merger formation 
\citep{Belczynski2007,Mennekens2014}; a finding that was not encountered in previous studies. 
Only later was it noted that low metallicity stars will dominate the 
formation of massive BHs \citep{Belczynski2010b} and BH-BH mergers in general
\citep{Belczynski2010a}. Full scale predictions that took into account low 
metallicity stars were performed in advance of  
the 2015 LIGO detections \citep{Dominik2012,Dominik2013,Dominik2015,Rodriguez2015,Marchant2016,
Belczynski2016a,Mandel2016}.

Additionally, Population III stars have also been considered as possible venues for BH-BH 
formation and gravitational wave detection for over thirty years 
\citep{Bond1984,Belczynski2004,Kinugawa2014}. 
Finally, prior to 
the first BH-BH merger detections, Population I/II very massive stars ($>150\msun$; \citep{Crowther2010}) were also  
introduced into predictions of BH-BH merger rates \citep{Belczynski2014,Marchant2016}.  

Since the first detection of GW150914, several investigations have examined how double 
black hole binaries could have been produced from the evolution of massive stars, whether from 
classical isolated evolution in low-metallicity environments \citep{Belczynski2016b,Eldridge2016}; 
via the aid of rapid rotation and hence homogeneous chemical evolution \citep{deMink2016,
Woosley2016};  via Population III stars \citep{Hartwig2016,Inayoshi2016,Dvorkin2016}; or 
from dynamical formation in interacting environments \citep{Mapelli2016,Rodriguez2016}. 
Other more exotic scenarios have been introduced and discussed in the context of GW150914; 
dark matter primordial BH-BH formation \citep{Sasaki2016,Eroshenko2016}, formation of a BH-BH 
merger from a divided core of a massive rapidly rotating single star \citep{Loeb2016}, 
or formation of BH-BH mergers with disks around BHs formed from fallback material in weak 
supernova explosions \citep{Perna2016}. 
   
In this study we consider the effects of Pair-instability Pulsation Supernovae (PPSN) and
Pair-instability Supernovae (PSN) on BH-BH mergers. PPSN are associated with severe mass loss 
\citep{Heger2002,Woosley2007} that may significantly reduce BH mass and thus detectability 
of BH-BH mergers. PSN are expected to completely disrupt massive stars with no resulting BH formation 
\citep{Bond1984,Fryer2001,Chatzopoulos2012a} and thus suppress formation of BH-BH mergers. 
While PSN are taken into account in some of the predictions for BH-BH merger formation 
(e.g., \cite{Marchant2016,Mandel2016,Spera2016}), PPSN and associated mass loss have thus 
far been ignored in studies of BH-BH formation (e.g., \cite{Dominik2015,
Rodriguez2015,Belczynski2016a,Marchant2016,Mandel2016,Rodriguez2016,deMink2016,
Belczynski2016b,Eldridge2016}) with the exception of recent work by \cite{Woosley2016}. 
We quantify the effect of PPSN and PSN on BH-BH mergers in our isolated classical binary
evolution channel. 
In brief, these processes introduce a maximum mass of stellar-origin black holes which 
differs from previous projections; compare to, e.g., the review in \cite{AstroPaper}.

\section{Pair-instability: pulsation supernovae and supernovae}
\label{ex1}

Pair-instability pulsation supernovae and pair-instability supernovae are produced 
in massive stellar cores when the temperature in the inner part of the star becomes 
sufficiently high ($T\gtrsim 5\times 10^9\,$K) while the density is still rather low 
(e.g., oxygen burning) to allow for the production of electron--positron pairs. 
For not very massive stellar cores such high temperatures may also be 
reached during the advanced burning stages, however, then the entropy is lower 
and the density is higher, blocking the formation of electron--positron pairs. 
The pair production reduces the photon pressure by converting internal energy 
into rest mass of the pairs and effectively lowers the equation-of-state 
$\gamma$ in the radiation dominated plasma to unstable values ($\gamma<4/3$).  
This leads to rapid contraction with a small fraction of free fall 
acceleration of the inner part of the star, typically within the C/O core.   
The increase of temperature that follows the collapse leads to very rapid 
thermonuclear burning that can release energy comparable to, or even in 
excess of, the binding energy of the star. If enough energy is deposited
and a sufficient fraction of the star leaves the regime of pair instability, 
the contraction can evolve into an expansion that steepens into a
shock at the edge of the carbon-oxygen (CO) core.
Depending on the amount of deposited energy and momentum transfer, the outer 
layers may be ejected (pair instability pulsation supernova: PPSN) or the entire 
star is disrupted (pair instability supernova: PSN). In the 
latter case no remnant is left behind. In the former case, even after the first 
pulse the star may still be massive enough for subsequent and recurrent pulses. 

These pair-instability pulsation supernovae and associated mass ejections operate for 
helium cores with masses $M_{\rm He} \gtrsim 40$--$45\msun$. For helium cores above 
$M_{\rm He} \gtrsim 60$--$65\msun$ the (non-pulsational) pair instability operates, 
i.e., the first pulse is energetic enough to entirely disrupt the star \citep{Heger2002}. 
For even more massive helium cores $M_{\rm He} \gtrsim 135\msun$
the center of the star becomes so hot, that besides burning all the way to 
nickel, photo-disintegration of heavy elements to alpha particles and then subsequent 
photo-disintegration of alpha particles decreases photon pressure. This softens the 
equation of state enough that instead of explosion a collapse of the central region 
leads to the formation of a black hole. The release of thermonuclear energy is not 
able to overcome the large binding energy of the massive star \citep{Fryer2001,Heger2002}. 

These limits are subject to various modeling uncertainties. For example, the most 
recent calculations (Woosley et al., in preparation) show that pulses may appear 
already for helium cores with masses as low as $M_{\rm He} \sim 30$-$40\msun$. 
However, these pulsations will not remove significant mass from stars 
($M_{\rm ejecta}\lesssim1\msun$). The division between PPSN and PSN is also
not clear cut, and helium cores within the mass range $M_{\rm He} \sim 60$--$65\msun$ 
may either experience PPSN or PSN. 

The pair-instability pulsation supernovae for helium cores with $M_{\rm He} \sim 45$-$65\msun$ 
can remove significant amounts of mass from stars prior to the core collapse 
($M_{\rm ejecta} \sim 5$--$20\msun$). Beyond model assumptions, the exact amount 
of mass loss is sensitive to the detailed sequence of pulses and their energies 
\citep{Woosley2015} and therefore may vary discontinuously as a function of initial mass.  
In general, however, it tends to increase with the helium core mass. 
Despite the PPSN mass loss, these helium cores are massive enough to form black
holes after the final core collapse. The maximum helium star mass after PPSN is 
likely around $M_{\rm He}\sim 45 \pm 5 \msun$. For a given helium star mass at the end of core 
helium burning the final outcome will be insensitive to the evolutionary path to this 
point and insensitive to the initial metallicity of the progenitor \citep{Woosley2016}.

The above discussion fully applies to the evolution of massive close binaries in which 
stars get stripped of their H-rich envelopes during Roche Lobe Overflow (RLOF) or common 
envelope (CE) events independent of their mass or metallicity. However, if stars retain 
their H-rich envelope (e.g., low metallicity single stars) the pair-instability
pulsation supernovae models allow for stars to have masses as high as $M \sim 52\msun$ at 
the time of core collapse (Woosley et al., in preparation). The same models give the absolute 
upper limit on the star mass at the time of core collapse in the mass regime in which PPSNe 
operate: $M \sim 70\msun$. These values are based on a non-rotating stellar model with no 
wind mass loss (e.g., POP III star) and with almost no pair-instability pulsation supernovae  
mass loss ($M_{\rm He} \sim 30\msun$). This is an important difference from helium star models 
(binary evolution) that provide only up to $\sim 45 \msun$ for black hole formation, 
while models that retain H-rich envelopes (single stars) can potentially supply up to 
$70\msun$ to form a black hole. 

Stars that retain their hydrogen envelope are expected to form black holes with mass 
up to $\sim 50$--$70\msun$, and then above $\sim 135\msun$ (if massive stars reach conditions 
sufficient for the formation of helium cores above $135\msun$). Stars that become 
naked helium cores during the evolution are expected to form black holes with mass 
up to $\sim 45 \pm 5\msun$, and then above $\sim 135\msun$. 
This gap in the mass spectrum of BHs: $\sim 50$--$135\msun$ is the second mass gap 
predicted for compact object formation \citep{Yusof2013,Belczynski2014,Marchant2016}. 
The first mass gap appears between the heaviest NSs and the lightest BHs: 
$\sim 2$--$5\msun$ \citep{Bailyn1998,Ozel2010}. It has been explained either in 
terms of observational bias on BH mass measurements \citep{Kreidberg2012},  
or in the context of the supernova explosion engine \citep{Belczynski2012}, or in  
terms of the transition of compressible nuclear matter into incompressible 
quark-superfluid changing heavy NSs into dark energy (invisible) objects 
\citep{Hujeirat2016}. 

Most of the work studying pair-instabilities has focused on $1$-dimensional explosions 
\citep{Barkat1967,Bond1984,Woosley2007,Heger2002,Chatzopoulos2012a,Chatzopoulos2012b}. 
Given the turbulent nature of this runaway burning, it is not surprising that there 
are some differences between these $1$-dimensional models and recent $2$-dimensional 
results \citep{Chatzopoulos2013,Chen2014}. However, these differences are small and 
insignificant for the purpose of this paper. Much more work remains to be done to 
understand the exact nature of these explosive events. Observations of BH-BH mergers 
with heavy black holes will impose very useful constraints on pair-instability models.

\section{Model}
\label{model}

Population synthesis calculations were performed with the {\tt StarTrack} code 
\citep{Belczynski2002,Belczynski2008a}. Recently we updated this code with improved
physics. The improvements relevant for massive star evolution include updates to the
treatment of CE evolution \citep{Dominik2012}, the compact object masses
produced by core collapse/supernovae \citep{Fryer2012,Belczynski2012}, stellar binary
initial conditions set by observations \citep{deMink2015}, and observationally
constrained star formation and metallicity evolution over cosmic time
\citep{Belczynski2016b}.

In particular, our calculations employ analytic fits to non-rotating stellar models 
presented by \citet{Hurley2000}. More advanced stellar models are now available 
\citep[e.g.][]{Chen2015} and some include effects of rotation on stellar evolution  
\citep{Georgy2013,Choi2016}. We do not directly use any of these updated stellar  
models. However, we updated the original \citet{Hurley2000} models with revised stellar 
wind prescriptions \citep{Vink2011}, and with a new compact object formation scheme 
\citep{Fryer2012}, and we have begun calibrating our evolution with the calculations 
performed with these most modern stellar models \citep{Pavlovskii2016}.
Treatment of convection, mixing, stellar rotation and winds differs from code to code 
resulting in different stellar core mass for the same star. These uncertainties impose 
limitations to any study performed with results of stellar evolutionary calculations. 
In particular, we use the helium core mass to judge the development of pair-instability  
and carbon-oxygen core mass to calculate neutron star and black hole mass (see below). 
To quantify uncertainties that originate from details of stellar modeling it would be
necessary to repeat our calculations with a different population synthesis code that employs
different stellar models than used here. Such codes already exist 
\citet{Spera2015,Marchant2016,Eldridge2016} and hopefully they will be used in the near
future to test and challenge our current predictions. 

Initial parameters for massive binary stars (progenitors of neutron stars
and black holes; $M_{\rm ZAMS} \gtrsim 7$--$10 \msun$) are guided by recent
observations of O/B binaries \citep{Sana2012,Kobulnicky2014}. The primary mass is 
chosen from a three component broken power-law 
initial mass function with a rather flat power-law exponent for massive stars 
($\alpha_{\rm IMF}=-2.3$), a flat mass ratio distribution is used to calculate
the secondary mass, binaries are assumed to form predominantly on close 
($\propto \left(\log P_{\rm orb}\right)^{-0.5}$) and rather circular orbits 
($\propto e^{-0.42}$). We assume that binarity for massive stars ($>10\msun$) 
is $100\%$ and $50\%$ for less massive stars. Our stars are assumed to only
moderately rotate ($200$--$300$ km s$^{-1}$). We do not consider the small fraction 
of massive stars that may rotate at very high speeds ($\sim 600$ km s$^{-1}$). 
For such rapidly rotating stars the effects of rotation on their evolution
need to be included in evolutionary calculations \citep{Marchant2016,deMink2016,Eldridge2016}. 
As reported, the stellar models used in our study do not include effects of
rotation on the evolution of stars. However, we include the effect of rotation in
estimates of tidal interactions between both stars and the binary orbit. 
Stellar spins (and thus binary orbit) are further affected due to magnetic
braking for stars with significant convective envelopes. Additionally, we
calculate orbital changes due to stellar wind mass loss (Jeans mode mass
loss: orbital expansion) and angular momentum loss due to emission of
gravitational waves (important only for very compact binaries). 
The development and (dynamical) stability of RLOF, is judged based on: binary  
mass ratio, evolutionary stage of the donor, response to mass loss, and behavior 
of the orbital separation in response to mass transfer and angular momentum
loss~\citep{Belczynski2008a}. 
During stable RLOF we assume that half of the mass is accreted onto the companion, 
while the other half is lost from the binary with the specific angular momentum; 
we adopt rather effective angular momentum loss with $j_{\rm loss}=1.0$ defined in
~\citet{Podsiadlowski1992}. Unstable mass transfer is assumed to lead 
to CE. CE is treated using the energy balance formalism with fully effective 
conversion of orbital energy into envelope ejection ($\alpha=1.0$), while the 
envelope binding energy is obtained with the parameter $\lambda$ which depends on 
donor mass, radius, and metallicity. A typical value of this parameter is: 
$\lambda \approx 0.1$ for massive stars~\cite{Xu2010}. During CE neutron stars 
and black holes accrete at $10\%$ Bondi-Hoyle rate~\citep{Ricker2008,Macleod2015}.
Supernova explosions affect the binary orbit; we allow for mass loss, neutrino loss 
and natal kicks during explosions. In particular, for specific configurations a 
given binary may be disrupted and two stars are then evolved in isolation. 

In this study we add to the {\tt StarTrack} code a model for mass loss associated with
pair-instability pulsation supernovae, and we incorporate new input physics that allow 
for pair-instability supernovae. We will refer to this input physics as the M10 model
(based on the discussion presented in ~\ref{ex1}). 
Our standard input physics without PPSN/PSN will be referred to as model M1 (as in 
\cite{Belczynski2016b}). Our basic assumptions in compact object mass calculations apply 
to both models: M1 and M10. However, in model M10 we impose one extra constraint on 
BH mass. The details for both models are given below. 

Each compact object mass is estimated based on a selection of hydrodynamical supernova 
models that are initiated with stellar models obtained with detailed evolutionary 
calculations~\citep{Woosley2002,Heger2003,Fryer2006,Limongi2006,Dessart2007,
Poelarends2008,Young2009}. The final prescriptions for compact object mass are
based on the mass of a star at the time of core collapse and the final mass of its 
carbon-oxygen core~\citep{Fryer2012}. In this study we employ a prescription that 
reproduces the observed mass gap between neutron stars and black holes with rapid 
supernova explosions~\citep{Belczynski2012}. Neutron stars are formed with
masses in the range $1.1$--$2.5 \msun$, while black holes form with masses in the range
$5$--$94\msun$ and the upper limit on a black hole mass is set by stellar
winds and their dependence on metallicity~\citep{Belczynski2010b}. For single stars 
our formulation results in a strict limit between neutron stars and black holes in 
terms of the Zero Age Main Sequence mass ($M_{\rm ZAMS}$) of the compact object
progenitor. For solar metallicity ($Z=0.02$) this limit is found at 
$M_{\rm ZAMS}\approx20\msun$ and it changes with metallicity of the progenitor.
This prescription has been designed to reproduce observations of compact 
objects. More pertinently, our standard model with and without pair instability
supernovae reproduces the masses and event rates of LIGO observations (see
Sec.~\ref{ex3}).

Some recent hydrodynamical simulations of core collapse seem to indicate that even for 
single stars (initial mass range $M_{\rm ZAMS}\approx15$--$40\msun$) one may expect 
non-monotonic behavior in neutron star and black hole formation~\citep{Oconnor2011,Ertl2016}. 
These conclusions reflect violent, chaotic behavior that impacts the final collapsing 
core and hence remnant mass. Although \cite{Fryer2012} found similar behavior, the exact 
results depend sensitively on the stellar evolution code used to produce the progenitor. 
Our prescription uses the smoothed fit from \cite{Fryer2012}. \cite{Fryer2014} provides a 
review of the different methods to determine compact remnant masses and their similarities.

Introducing a chaotic single star initial-remnant mass relationship would not 
influence our predictions for the most massive stars, including the focus of this 
work (pair-instability pulsations supernovae) and our predictions for events like 
GW150914 ($36+29 \msun$ BH-BH merger). On the other hand, however, adding a chaotic 
mechanism would decrease the number of lower-mass BH-BH mergers formed from stars with 
masses $M_{\rm ZAMS}<40\msun$. In particular, the detection rate of BHs with low mass 
like GW151226 ($14+8\msun$ BH-BH merger) would be reduced.

Finally, we note that our model already incorporates nondeterministic elements in 
binary evolution that create comparable effects to chaos and significantly impacts 
the range of initial star masses that can form BHs. For example, even stars as massive 
as $100\msun$ may form NSs in some specific binary configurations, while in other 
binary configurations they will form BHs~\citep{Belczynski2008c}. In other words, our 
simulations include non-monotonic formation of NSs and BHs with respect to progenitor 
initial mass, with NSs and BHs mixed up in wide range of initial masses 
($M_{\rm ZAMS}\approx10$--$100\msun$) in our binary star simulations.

In model M10 we assume that stars that form helium cores in the mass range 
$M_{\rm He}=45$--$65\msun$ are subject to PPSN and they lose \emph{all} the mass above 
the central $45\msun$ just prior to core collapse. This includes the entire H-rich envelope 
(if a given star has retained one), or it means losing outer He-rich and possibly 
C/O-rich layers from a Wolf-Rayet star. The remaining star is then assumed
to directly collapse to a BH \citep{Fryer2012}. We only allow for mass loss
via neutrino emission:  
\begin{equation}
M_{\rm BH} = 45.0\ (1.0-f_{\rm n})\ \msun. 
\label{bhm}
\end{equation}
The BH mass is mostly set by the PPSN threshold, however it should be
remembered that this threshold is subject to a number of modeling uncertainties  
and most likely it is not a sharp cutoff but rather a transition region centered 
around this threshold value ($M_{\rm He} \approx 45 \pm 5\msun$; see Sec.~\ref{ex1}).
We do not introduce this PPSN threshold uncertainty on our eq.~\ref{bhm}, but we 
take it into account while reporting our final results (e.g., maximum mass from the 
above equation is $50\msun$ if this uncertainty are taken into account). 
The amount of neutrinos that can escape ($f_{\rm n}$) during massive BH formation is 
highly uncertain. Neutrino mass losses during NS formation are at the level of 
$f_{\rm n}=0.1$~\citep{Lattimer1989,Timmes1996}. However, one may expect that during 
BH formation some neutrinos are trapped under the BH horizon $f_{\rm n}<0.1$. In this 
study we adopt $f_{\rm n}=0.1$ neutrino mass loss for BH formation. This generates 
$M_{\rm BH}=40.5\msun$ from stars subjected to PPSN mass loss. Had we assumed 
$f_{\rm n}=0.01$ that would result in $M_{\rm BH}=44.6\msun$ from stars subjected to 
PPSN mass loss. Note that we additionally assume that the fraction of BH mass lost 
in neutrinos is independent of BH mass. 

The mass of a BH given by eq.~\ref{bhm} indicates maximum BH mass for BH
formation out of naked helium stars. For no neutrino mass loss and the high-end
of PPSN threshold it is $M_{\rm BH} = 50 \msun$. This applies to the formation of 
BH-BH mergers in isolation whether via the classical or homogeneous evolution channel.  
It is possible that at low metallicity ($\lesssim 1\% \zsun$) a {\em single} star 
with a helium core mass just below the PPSN threshold retains some of its H-rich 
envelope. In such a case, the BH can reach a mass higher than given by eq.~\ref{bhm}
(see Sec.~\ref{ex2} for details). 

Also in model M10 we assume that stars that form helium cores in the mass range 
$M_{\rm He}=65$--$135\msun$ are subject to PSN. In such cases we assume that the entire 
star is disrupted and no remnant forms (neither NS nor BH). 

We also consider an extra model that differs from model M1 by only one parameter. 
In model M3 we subject all compact objects (both neutron stars and black holes) to 
high natal kicks. These kicks are adopted from the natal kick distribution measured for 
single pulsars in our Galaxy \citep{Hobbs2005}; the natal kicks are drawn from a $1$-D 
Maxwellian with $\sigma=265$ km s$^{-1}$ (average $3$-D speed of $\sim 400$ km s$^{-1}$).
In other models (M1 and M10), compact objects that experience fall back during the 
formation receive a decreased kick (inversely proportional to the amount of fall 
back). For the most massive BHs (full fall back) there is no natal kick. But even heavy NSs 
receive reduced natal kicks as there is some fall back expected at their formation 
\citep{Fryer2012}. This assumption on compact object natal kicks leads to a severe 
reduction of BH-BH merger rates, moderate reduction of BH-NS merger rates, and small 
reduction of NS-NS merger rates; we consider it our \emph{pessimistic} model 
\citep{Belczynski2016b}.  

Our population synthesis data are set in a cosmic framework. The merger rate
densities of double compact objects are obtained as a function of redshift,
and the merger properties as a function of type of merger, its mass, and its 
origin \citep{Belczynski2016a}. 

\begin{figure}
   \hspace*{-0.6cm}
   \includegraphics[width=10.5cm]{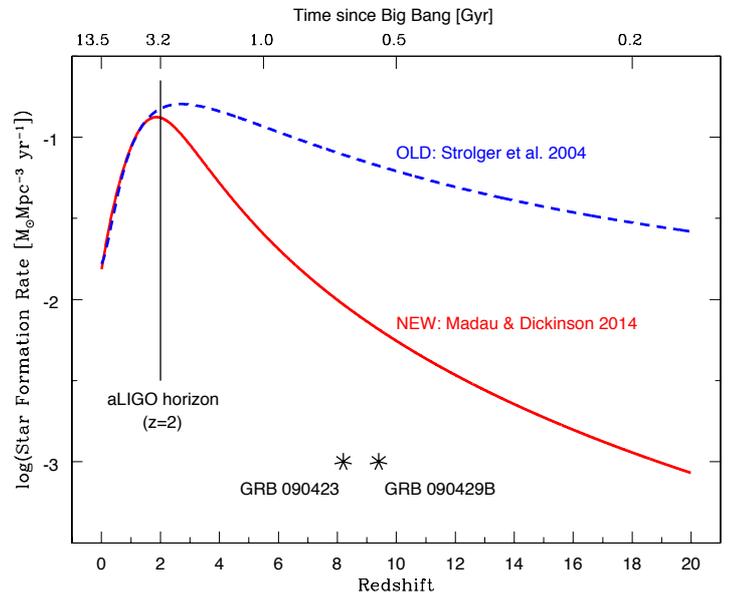}
   \vspace*{-2.8cm}
   \caption{
    Star formation rate (SFR) as a function of redshift and cosmic time. The
    blue dashed line shows the SFR used in our previous studies, while the red line
    indicates the most recent update (used in this study). Note that for high 
    redshifts ($z>2$) the updated SFR generates less stars, and thus reduces 
    formation rates of BH-BH progenitors. We mark the two highest redshift GRBs. 
    Since GRBs form from stars or stellar remnant mergers, their positions mark 
    the lower limit on the beginning of star formation. We also mark the detection 
    horizon for the full advanced LIGO design sensitivity. We assume that 
    Population II and I stars (considered in this study) form at redshifts 
    $z<15$.
   }
   \label{fig.sfr}
   \end{figure}

For the cosmic star formation rate (SFR) we adopt a formula from the recent study of 
\citet{Madau2014}: 
\begin{equation}
\mbox{SFR}(z)=0.015 {(1+z)^{2.7} \over 1+[(1+z)/2.9]^{5.6}} \,\msun \, {\rm
Mpc}^{-3}\, {\rm yr}^{-1}.
\label{sfr}
\end{equation}
SFR is well established at low redshifts ($z<2$), however there is a lot of 
uncertainty at higher redshifts (reddening and scarcity of good observational
constraints). Our adopted SFR results in very low star formation rates for $z>2$. 
This formula may be treated as a proxy for the lower bound on SFR at high redshifts. 
It is possible that the type of analysis used by \citet{Madau2014} does not fully 
correct for the small galaxies not seen in UV surveys; the predicted high-redshift 
SFR cannot reionize the Universe with stars; and it underpredicts the observed GRB 
rate \citep{Kistler2009,Horiuchi2011,Mitchell2015}.   
Any increase of SFR would result in an increase of our double compact object merger rates. 
In Figure~\ref{fig.sfr} we show~\citet{Madau2014} the SFR adopted in our study in 
comparison with the very high SFR from a different study~\citep{Strolger2004}. Most 
likely, the actual high-redshift SFR would be found somewhere between these two 
models. 

We adopt the mean metallicity evolution model from ~\citet{Madau2014}, and we modify
it to: 
\begin{equation}
\log(Z_{\rm mean}(z))=0.5+\log \left(  {y \, (1-R) \over \rho_{\rm b}}
\int_z^{20} {97.8\times10^{10} \, sfr(z') \over H_0 \, E(z') \, (1+z')} dz'
\right)
\label{Zmean}
\end{equation}
with a return fraction of $R=0.27$ (mass fraction of stars put back into the 
interstellar medium), a net metal yield of $y=0.019$ (mass of metals ejected into 
the medium by stars per unit mass locked in stars), a baryon density 
$\rho_{\rm b}=2.77 \times 10^{11} \,\Omega_{\rm b}\,h_0^2\,\msun\,{\rm Mpc}^{-3}$ 
with $\Omega_{\rm b}=0.045$ and $h_0=0.7$, a SFR from eq.~\ref{sfr}, and
$E(z)=\sqrt{\Omega_{\rm M}(1+z)^3+\Omega_{\rm k}(1+z)^2+\Omega_\Lambda)}$
with $\Omega_\Lambda=0.7$, $\Omega_{\rm M}=0.3$, $\Omega_{\rm k}=0$,
and $H_0=70.0\,{\rm km}\,{\rm s}^{-1}\,{\rm Mpc}^{-1}$.
In our modification we have increased the mean level of metallicity by $0.5$
dex to be in a better agreement with observational data~\citep{Vangioni2015}. 
We assume a log--normal distribution of metallicity around the mean at each 
redshift, and we adopt $\sigma=0.5$ dex from~\citet{Dvorkin2015}. 
The graphic presentation of our adopted model is given in {\em Extended Data
Figure 6} of \citet{Belczynski2016b}. The formation of BH-BH mergers is very
sensitive to metallicity \citep{Belczynski2010a,Dominik2013,deMink2015}; the
rise/drop of the mean level of metallicity will cause the decrease/increase in 
BH-BH merger rates, respectively.
As more stringent constraints appear on SFR and metallicity evolution we will
incorporate them into our modeling and test the influence of associated
uncertainties on double compact object merger rates. 

For each merger we model the full inspiral-merger-ringdown waveform using the 
IMRPhenomD gravitational waveform templates \citep{Khan2015,Husa2015}. We
have repeated our calculations with IMRPhenomC waveforms and detection rates
of double compact objects are within $10\%$ of these obtained with the IMRPhenomD 
waveform family. A merger is considered a detection if the signal-to-noise
ratio in a single detector is above a threshold equal to $8$. This is a simple 
proxy for detectability by a detector network.
We estimate detection rates as described in \citet{Belczynski2016a}. For
increased accuracy with respect to \citet{Belczynski2016a}, where we used an
analytic fit [Eq.~12 of \citet{Dominik2015}] to the cumulative distribution
function describing the detector response, here we interpolate the numerical
data for the cumulative distribution function available online at
\url{http://www.phy.olemiss.edu/~berti/research.html}. This improvement
leads to a small increase ($\sim$ few per cent) of detection rates with
respect to previous work \citep{Belczynski2016a,Belczynski2016b}.

\section{Mass of single BHs and BH-BH mergers}
\label{ex2}

   \begin{figure}
   \hspace*{-0.6cm}
   \includegraphics[width=10.5cm]{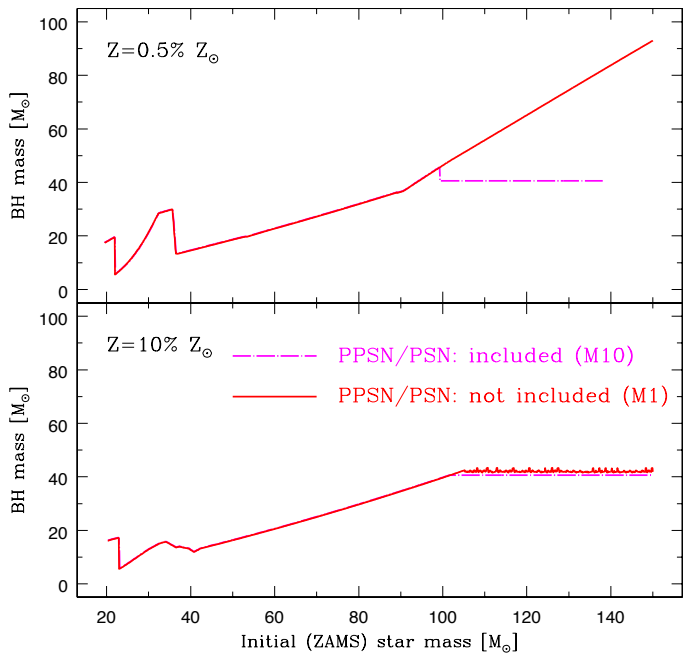}
   \vspace*{-2.8cm}
   \caption{
    Initial---final mass relation for single stars. 
    Models with (M10) and without (M1) pair-instability pulsation supernovae and 
    pair-instability supernovae are shown.
    \emph{Bottom panel:} At high metallicity ($Z=10\% \zsun$ and higher) 
    the models are indistinguishable. For the border-line metallicity of $Z=10\% \zsun$ 
    stars with very high initial mass ($M_{\rm zams}>100\msun$) will form slightly 
    lighter BHs (by $\sim 2 \msun$) if pair-instability pulsation supernovae are included. 
    \emph{Top panel:} At low metallicity (e.g., $Z=0.5\% \zsun$) pair-instability 
    pulsation supernovae and pair-instability supernovae do not allow for high mass 
    BH formation; maximum BH mass is $M_{\rm BH}=45\msun$. For the very low metallicity 
    of $Z=0.5\% \zsun$, very massive stars ($M_{\rm zams}\approx 100$--$140\msun$) 
    lose significant mass in pair-instability pulsation supernovae reducing the BH mass 
    to $M_{\rm BH}\approx40\msun$, while the most massive stars ($M_{\rm zams}>140\msun$) 
    explode in pair-instability supernovae leaving no remnant. 
   }
    \label{bhmass}
    \end{figure}

   \begin{figure}
   \hspace*{-0.6cm}
   \includegraphics[width=10.5cm]{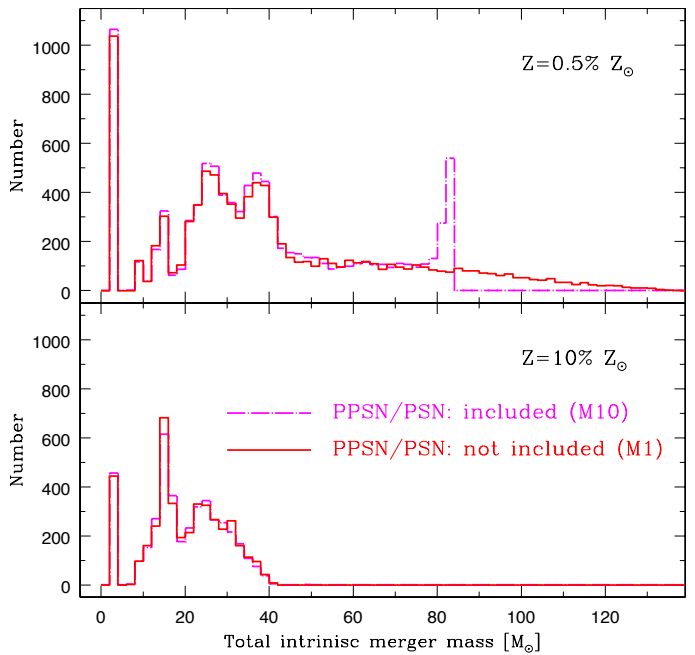}
   \vspace*{-2.8cm}
   \caption{
    Total intrinsic (not weighted by merger rate nor by detection probability) merger 
    mass distribution for two progenitor stellar populations of different metallicity.   
    Models with (M10) and without (M1) pair-instability pulsation supernovae and
    pair-instability supernovae are shown. 
    \emph{Bottom panel:} At high metallicity ($Z=10\% \zsun$ and higher) 
    models are indistinguishable. 
    \emph{Top panel:} At low metallicity (e.g., $Z=0.5\% \zsun$) pair-instability 
    pulsation supernovae and pair-instability supernovae do not allow for high mass 
    merger formation ($M_{\rm tot}\lesssim 80\msun$). Note that the model that does 
    not take into account pair-instability pulsation supernovae and pair-instability 
    supernovae allows for the formation of high mass mergers ($M_{\rm tot} > 80\msun$).
   }
    \label{bhbhmass}
    \end{figure}

The mass spectrum of single BHs remains unaffected by PPSN/PSN for progenitor stars 
with metallicity higher than $Z>10\%\zsun$. For solar metallicity we adopt $\zsun=0.02$ 
\citep{Villante2014}. This is the result of the fact that stars at high 
metallicity are subject to intensive stellar wind mass loss \citep{Vink2011} 
and they do not form helium cores above $45\msun$ (see Fig.~\ref{bhmass}).

However, in the metallicity range $Z=1$--$10\%\zsun$, the mass spectrum of BHs remains 
unaffected by PSN but \emph{is} affected by PPSN. 
Stars that form within the initial mass range $M_{\rm zams}\approx 100$--$150\msun$ form 
BHs with the upper limit of their mass set by PPSN mass loss: $M_{\rm BH}=40.5\msun$ 
(see Fig.~\ref{bhmass}). 

For the lowest metallicities considered in our study, $Z=0.5$--$1\%\zsun$,
the most massive stars are affected by \emph{both} by PPSN \emph{and} PSN. 
Stars in the mass range $M_{\rm zams}\approx 20$--$100\msun$ are not affected by
PPSN nor by PSN and they form a wide range of BH masses. The highest mass of a BH is 
$M_{\rm BH }=45\msun$ and it is formed by a star with initial mass 
$M_{\rm zams}\approx 100\msun$ that at time of core collapse has a total mass of 
$50\msun$; $5\msun$ of H-rich envelope and $45\msun$ core (with the top $10\msun$
being He-rich, while the deeper layers consist of heavier elements). If neutrino mass 
loss in core collapse is not as effective as we have assumed ($10\%$ mass loss) then 
the maximum BH mass would be $M_{\rm BH }=50\msun$ ($0\%$ neutrino mass loss).  
Stars in the initial mass range $M_{\rm zams}\approx 100$--$140\msun$ form BHs with 
mass set by PPSN mass loss: $M_{\rm BH }=40.5\msun$, while stars in the mass range 
$M_{\rm zams}\approx 140$--$150\msun$ are disrupted by PSN and they leave
no remnant (see Fig.~\ref{bhmass}).

Here we encounter an important caveat. The maximum mass of a BH formed by a single 
star in our simulations is $50\msun$ as reported above. However our simulations are 
performed for a limited metallicity range ($Z=0.03$--$0.0001$ or $Z=0.5\%$--$150\%\zsun$).   
Had we extended the metallicity range down to Population III stars ($Z \approx 0$) 
then the maximum BH mass from single stellar evolution could be higher. Stellar winds 
for Population III stars are expected to be very weak and these stars may retain most 
of their H-rich envelope. If our $M_{\rm zams}\approx 100\msun$ model retained 
the {\em entire} H-rich envelope and if it still had formed an He core {\em below} PPSN threshold, 
then this star could have potentially formed a $100\msun$ BH (no neutrino loss and no 
supernova mass loss). Most likely the mass of the He core of such star would be above the PPSN 
threshold due to increased central temperature, but this sets an upper limit on the 
maximum BH mass for single stars. If Population III stars rotate rapidly (no angular 
momentum loss with stellar winds) then they form cores that are more massive than 
predicted in our simulations of Population I and II stars. Therefore, the maximum BH mass 
for single stars of very low metallicity ($Z<0.0001$) is expected somewhere in the range 
$50$--$100\msun$. In the case of rapid rotation, when the entire star is transformed into 
an He-rich object at the end of the main sequence (homogeneous evolution) the maximum BH mass is 
$\sim 50\msun$ even for very low/zero metallicity. For slow rotators, the maximum BH mass 
is most likely closer to $\sim 100\msun$.  
This rather complex picture is simplified in the case of binary evolution leading to the 
formation of BH-BH mergers in isolation (no dynamical interactions). In the case of classical 
evolution performed in this study, the formation of massive BH-BH mergers is {\em always} 
preceded by both stars being stripped of their H-rich envelope during progenitor
binary evolution~\citep{Belczynski2016b}.
In the case of homogeneous evolution (also field binaries), the progenitor stars burn all 
the H-rich envelope into an He core~\citep{Marchant2016,deMink2016}. In both cases, the maximum 
(individual) BH mass in the BH-BH merger is set by eq.~\ref{bhm}, and depending on the assumption on 
neutrino losses it is found at $\sim 40$--$45\msun$ (see below).   

Evolution of binary stars and associated intrinsic double compact object total 
merger mass is demonstrated in Figure~\ref{bhbhmass}. We define double compact object 
intrinsic total merger mass as the total binary mass: $M_{\rm tot}=M_{\rm 1}+M_{\rm 2}$ in 
the source frame where $M_{\rm 1}$ and $M_{\rm 2}$ are two compact object masses. 
This Figure shows BH-BH, BH-NS and NS-NS mergers formed out of a population of 
the same number of isolated massive binaries for two metallicities ($Z=10\%\zsun$ 
and $Z=1\%\zsun$) in our classical evolutionary scenario. Only binaries with a 
merger time shorter than the Hubble time ($13.7$ Gyr) are shown. 

At high metallicity $Z\gtrsim10\%\zsun$ the BH-BH merger mass is not affected by the 
new input physics. This is the result of high wind mass loss at high metallicity, and 
massive stars are not subjected to either PPSN or PSN, thus the mass spectrum is unchanged.

At lower metallicities, stars can undergo PPSN or PSN, with noticeable influence on the 
merger mass distribution. Due to PPSN, no black hole binaries with a total mass 
above $M_{\rm tot}\sim80\msun$ form or merge, producing a sharp cutoff in the mass 
distribution of coalescing BH-BH binaries. Moreover, due to PPSN, black holes which 
in M1 would have formed from helium cores with masses $45<M_{\rm he}<135\msun$ have 
in M10  a prescribed (and lower) final black hole mass. In particular, two massive 
stars subjected to PPSN will have a total merger mass of $M_{\rm tot}\sim80\msun$;  
and this produces a strong abundance of BH-BH mergers just below the cutoff.  
BHs that disappear due to PSN in M10, would have mass $M_{\rm BH}>80\msun$ in model M1. 
Since these BHs form only at very low metallicities ($Z\lesssim0.5\%\zsun$) and since 
they form only from very massive stars ($M_{\rm zams}>140\msun$) there are
so few of them that their impact is negligible on our predictions. 

We do not consider stars above $M_{\rm zams}>150\msun$ in this study. Such
massive stars exist \citep{Crowther2010} and were already considered in
terms of BH formation \citep{Yusof2013} and as progenitors of massive BH-BH
mergers \citep{Belczynski2014,Marchant2016}.

\section{BH-BH merger and detection rates}
\label{ex3}

   \begin{figure}
   \hspace*{-0.6cm}
   \includegraphics[width=9.9cm]{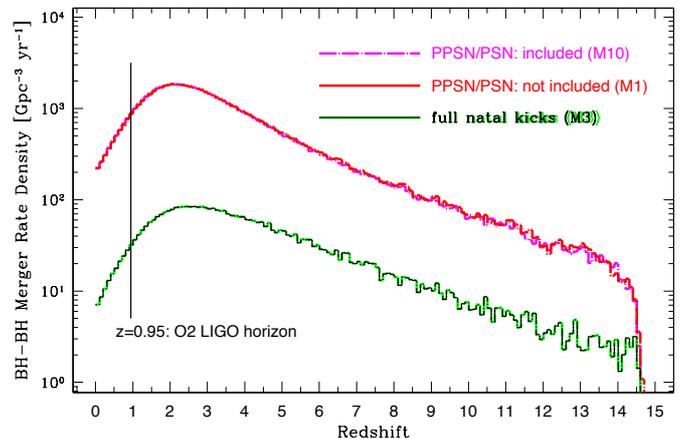}
   \vspace*{-2.8cm}
   \caption{
    Source frame merger rate density for BH-BH binaries. Note that there is virtually 
    no change of the rate at any redshift for models with (M10) and without (M1) 
    pair-instability pulsation supernovae and pair-instability supernovae. For reference, 
    we plot the horizon distance for our most massive BH-BH merger in model M10 
    ($M_{\rm tot} \approx 80\msun$; see Fig.~\ref{bhbhmass}). We also show our pessimistic 
    model with high natal kicks (M3). 
   }
    \label{bhbhrate}
    \end{figure}

   \begin{figure}
   \hspace*{-0.6cm}
   \includegraphics[width=9.9cm]{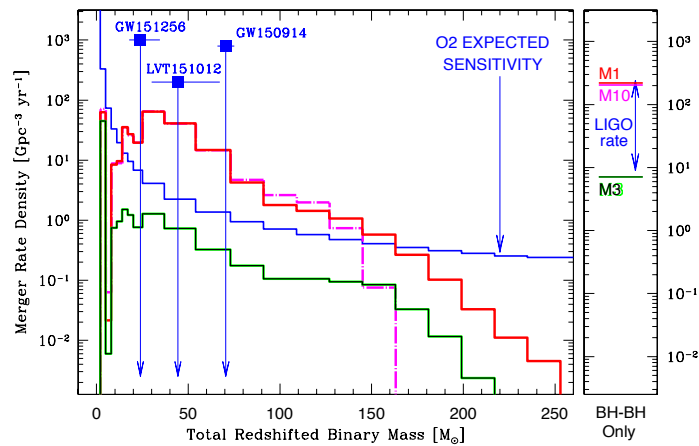}
   \vspace*{-2.8cm}
   \caption{
    \emph{Left panel:} Observer frame (redshifted) total merger mass distribution for 
    our models: with (M10; magenta broken line) and without (M1; red solid line) 
    pair-instability pulsation supernovae and pair-instability supernovae included. 
    The fiducial O2 LIGO upper limits are marked; the most likely detections are 
    expected when our models are above these upper limits. Note that both models are 
    hardly distinguishable by total merger mass with O2 observations, despite the fact 
    that more massive BH-BH mergers are produced in model M1. The two detections and the 
    next loudest gravitational-wave event from LIGO O1 observations \citep{LigoO1b} are marked: 
    GW150914 ($36+29\msun$ BH-BH merger at $z\approx0.1$), 
    GW151226 ($14+8\msun$ BH-BH merger at $z\approx0.1$), and  
    LVT151012 ($23+13\msun$ BH-BH merger at $z\approx0.2$). 
    \emph{Right panel:} Source frame BH-BH merger rate density for the local Universe. 
    The local source frame BH-BH merger rate density estimated by LIGO from
    the O1 run: $9-240\gpy$ (range marked with blue arrows) may be compared to our 
    local ($z<0.1$) source frame rate of $R_{\rm BHBH}\approx220\gpy$ (M1 and M10 models). 
    Both model rate densities are almost the same and are within the LIGO estimate. 
    We additionally show the results for our pessimistic model (M3) with high natal kicks: 
    $R_{\rm BHBH}\approx7\gpy$ (just below of the LIGO allowed range). 
    }
    \label{final}
    \end{figure}

The results and discussion in Section~\ref{ex2} indicate that PPSN play an important
role for stars at low metallicity, but no role for stars in near-solar-metallicity 
environments. To assess the overall impact of PPSN, we need to account for the 
heterogeneous and time-varying universe, by convolving the results of our previous 
analysis with a model for cosmic metal enrichment and star formation evolution.
Specifically, at each cosmic time we sample the evolution of stars with a broad spectrum 
of metallicities specific to this given time \citep{Belczynski2016b}. Double compact 
objects that are formed out of stars from each cosmic time are propagated forward in 
time (delay caused by stellar evolution and time to merger) and their merger times 
(and corresponding merger redshifts) are recorded. 

In Figure~\ref{bhbhrate} we show the BH-BH merger rate density as a function of redshift 
for our two models, one with (M10) and one without (M1) PPSN/PSN included. We note that both 
models are indistinguishable based on merger rate density alone. We also note that 
both models are consistent with the LIGO estimate of the local BH-BH merger rate 
density of $9$--$240\gpy$~\citep{LigoO1a,LigoO1b}. This LIGO estimate was based on the first two
detections (GW150914 and GW151226) and the next loudest event (LVT151012) found in the 
O1 data. Our local source frame BH-BH merger rate density is 
$R_{\rm BHBH}\approx 220\gpy$ ($z<0.1$), and $R_{\rm BHBH}\approx 250\gpy$ ($z<0.2$) for 
both models. The predicted BH-BH merger rate density first increases with redshift (by a 
factor of $\sim 10$ from $z=0$ to $z=2$), and then decreases with redshift. This behavior 
closely resembles the star formation rate history \citep{Madau2014}, and reflects the fact 
that the BH-BH merger delay time distribution follows a power-law $\propto t_{\rm delay}^{-1}$ 
\citep{Dominik2012,Belczynski2016b}. We also show (Fig.~\ref{bhbhrate}) our pessimistic 
model with high compact object natal kicks (M3). The local merger rate density is rather low: 
$R_{\rm BHBH}\approx 7\gpy$ ($z<0.1$). Within the framework of our model assumptions 
and simplifications (see Sec.~\ref{model}) this model is in tension with the LIGO 
estimate. This statement is subject to degeneracy with other thus far {\em untested} model 
parameters that could potentially increase the BH-BH merger rate density. For example, an increase 
in the SFR at high redshifts ($z>2$) with respect to our adopted model (which is hard to exclude 
due to rather weak observational constraints) could potentially bring the high kick model back into
agreement with the LIGO estimate. We plan to present a detailed study of this and other 
similar degeneracies when more stringent rate constraints appear from LIGO's next 
observation run (O2) in 2017. At the moment it seems that full natal kicks (adopted in M3; 
see Sec.~ref{model}) for black holes and heavy neutron stars are not supported
by the LIGO data. 

Figure~\ref{final} shows the total redshifted (observer frame) mass ($M_{\rm tot,z}=M_{\rm tot}(1+z)$) 
distribution of NS-NS/BH-NS/BH-BH mergers within the LIGO reach of the projected O2 
scientific run with a NS-NS average detection distance of $d_{\rm nsns}=120$ Mpc. 
The detection distance corresponds to the \emph{optimistic} O2 target sensitivity described 
by \cite{LVC2013}.
For comparison, O1 observations were sensitive only to $d_{\rm nsns}=70$ Mpc. We have assumed 
that the O2 run will last $6$ months, and will produce $65$ days of coincident data (duty 
cycle $p=0.36$ of two LIGO detectors observing simultaneously). We have adopted a fiducial 
O2 noise curve (``mid-high'') from~\cite{LVC2013}. We show both of our models and contrast 
them with the fiducial estimate of the sensitivity of the O2 run (O2 expected upper limits). 
In mass bins where our models are above the upper limits ($M_{\rm tot,z}=14$--$150\msun$) 
we predict the most likely detections, and detections are less likely in mass bins in which our 
models are significantly below the upper limits ($M_{\rm tot,z}<14\msun$: NS-NS mergers and 
most BH-NS mergers, and $M_{\rm tot,z}>150\msun$: the heaviest and most redshifted BH-BH mergers).
The most likely detections are expected in three mass bins that exceed O2 upper limits 
by the highest factors: $M_{\rm tot,z}=25$--$73\msun$. 

In Table~\ref{tab1} we list local (within redshift of $z<0.1$) merger rate densities, as 
well as predicted O2 detection rates ($R_{\rm O2}$ yr$^{-1}$). The detection rate is easily 
transformed into a number of expected detection events in the O2 observational run (e.g., 
assuming $65$ effective O2 observation days): $R_{\rm det}=(65/365) R_{\rm O2}$. We find 
that for our standard evolutionary model, whether or not we include
pair-instability pulsation supernovae (with the associated mass loss) and pair-instability 
supernovae (with the total disruption of BH progenitors), BH-BH mergers will dominate the 
gravitational wave detections. In particular, we expect about $\sim 60$ BH-BH merger 
detections in the O2 run for our standard evolutionary assumptions (about $1$ per day of 
coincident observations of two LIGO detectors). The prediction is significantly lower 
for our pessimistic model M3; only $\sim 2$ BH-BH merger detections in the entire O2 run. 
Since model M3 is already below the LIGO empirical estimate we expect more
detections than predicted in this model, and thus $\gtrsim 2$ detections. 

In the mass regime in which we predict detections, both models (M1 and M10) are
almost indistinguishable (Fig.~\ref{final}). Only at very high total BH-BH merger mass 
($M_{\rm tot,z}>150\msun$) are the two models visibly different; the model with PPSN/PSN 
(M10) does not extend to as high total merger mass as the model that does not include 
PPSN/PSN (M1). 
The mass range in which the two models differ significantly is not likely to be deeply 
probed with the LIGO O2 observations. The number of expected detections in the entire O2 run is 
$3.5$ and $1.5$ for a total redshifted merger mass of $M_{\rm tot,z}>145\msun$ and $>163\msun$, 
respectively for model M1, while it is $0.2$ and $0$ for $M_{\rm tot,z}>145\msun$ and 
$>163\msun$, respectively, for model M10. This is not a significant difference,
especially if the uncertainties on the maximum mass of a black hole are 
taken into account (see Sec.~\ref{ex1}). However, since the sensitivity of LIGO during the O2 
observations is already projected to be on the verge of distinguishing the two models, it 
seems likely that the fully-advanced design sensitivity will provide useful constraints on PPSN 
mass loss.

\begin{table}
\caption{Merger rate densities and detection rates for LIGO O2 run}             
\label{tab1}      
\centering          
\begin{tabular}{c| c c c}     
\hline\hline       
model & rate density\tablefootmark{a} & O2 rate\tablefootmark{b} & O2\tablefootmark{c}\\ 
merger type   & [Gpc$^{-3}$ yr$^{-1}$] & [yr$^{-1}$] & [65 days] \\ 
\hline          
   M1 &&&\\         
   NS-NS &  70.5 &   0.27 &  0.05 \\  
   BH-NS &  28.8 &   1.41 &  0.25 \\
   BH-BH &   222 &    371 &  66.0 \\
   $M_{\rm tot,z}>145\msun$\tablefootmark{d} & & & 3.80 \\
   $M_{\rm tot,z}>163\msun$\tablefootmark{d} & & & 1.77 \\
\hline
   M10 &&&\\
   NS-NS &  73.7 &   0.31 &  0.06 \\  
   BH-NS &  27.1 &   1.33 &  0.24 \\
   BH-BH &   219 &    363 &  64.6 \\
   $M_{\rm tot,z}>145\msun$ & & & 0.28 \\
   $M_{\rm tot,z}>163\msun$ & & & 0 \\
\hline                  
   M3 &&&\\
   NS-NS &  50.5 &   0.20 &  0.04 \\  
   BH-NS &  1.75 &   0.07 &  0.01 \\
   BH-BH &  7.06 &   13.7 &  2.44 \\
   $M_{\rm tot,z}>145\msun$ & & & 0.49 \\
   $M_{\rm tot,z}>163\msun$ & & & 0.19 \\
\hline                  

\end{tabular}
\tablefoot{\\
\tablefootmark{a}{Local merger rate density within redshift $z<0.1$.}\\
\tablefootmark{b}{Detection rate for LIGO O2 observational run.}\\
\tablefootmark{c}{Number of LIGO detections for effective observation time in O2.\\}
\tablefootmark{d}{The most massive mergers that can potentially distinguish the models.}
}
\end{table}

\section{Conclusions}

We have incorporated pair-instability pulsation supernovae and pair-instability 
supernovae into predictions of double compact object merger rates and masses
in context of near future LIGO observations. We find that;

   \begin{enumerate}
      
\item 
The mass of Population I/II stellar-origin black holes is limited to $50\msun$ by severe 
mass loss imposed by pair-instability pulsation supernovae (see Fig.~\ref{bhmass}). 
This may be contrasted with earlier predictions that the maximum mass of black holes 
can reach $80$--$130\msun$ in the evolution of Population I/II stars with modest 
initial masses: $M_{\rm zams}<150\msun$~\citep{Zampieri2009,Mapelli2009,Belczynski2010b,
Spera2015,Spera2016}. 

This conclusion applies to black holes formed below the second mass gap (no
compact objects in the mass range: $50$--$135\msun$; see Sec.~\ref{ex1}) imposed by 
pair-instability pulsation supernovae and pair-instability supernovae. 

If stars reach high enough mass to avoid disruption by pair-instability supernovae 
(i.e., if they can form helium cores above $135\msun$) then black holes with mass 
above $135\msun$ may form. Such massive black hole formation would require any 
combination of: very high star mass $>200$--$300\msun$ (whether it is initial mass 
or mass of a stellar merger), or very low metallicity (e.g., Population III stars), 
or very rapid rotation (homogeneous evolution).

\item 
We show that the introduction of pair-instability pulsation supernovae and the 
associated mass loss does not affect our predictions for detection of NS-NS, BH-NS 
and BH-BH mergers during the LIGO O2 observational run. In particular, our isolated 
binary classical evolution channel produces a similar number of detections for the 
O2 run whether or not pair-instability pulsation supernovae and pair-instability 
supernovae are included; $\sim 60$ BH-BH merger detections with a total 
redshifted mass in the range $10$--$150\msun$. Detections of BH-NS and NS-NS mergers 
originating from our classical isolated binary evolution model are not very likely 
in O2 (see Tab.~\ref{tab1}). 

We also note that the detection rates may be significantly smaller if pessimistic 
assumptions on binary evolution are adopted (i.e., model M3). To demonstrate this we 
have allowed for high black hole and neutron star natal kicks to obtain: only $\sim 2$ 
BH-BH merger detections in the entire O2 run. Since this model is just below the 
current LIGO empirical BH-BH merger rate estimate, it may serve as a lower limit on the 
number of predicted detections during O2. However, note that we use the optimistic 
target O2 sensitivity in all our predictions.

\item 
The detection of very massive BH-BH mergers ($M_{\rm tot,z}>150$--$200\msun$; see 
Fig.~\ref{final}) could distinguish between models with and without pair-instability
pulsation supernovae and pair-instability supernovae. However, our results argue that 
such a detection is unlikely. A detection of any binary with a BH mass 
$M_{\rm BH}>50\msun$ will rule out our adopted model for mass loss by pair-instability 
pulsation supernovae. Such an observation would require reconsideration of physics 
currently believed to be driving pair-instability pulsation supernovae and 
pair-instability supernovae (see Sec.~\ref{ex1}). 

An alternate solution for the detection of a massive BH ($M_{\rm BH}>50\msun$) 
is that the massive BH was formed through dynamical interactions. Any dynamical 
interaction that increases BH mass (either merger of two BHs, or rapid accretion 
onto a BH in tidal disruption event) can potentially accomplish this.  

For example, the merger of two lighter BHs (first burst of gravitational waves)
may form a massive BH. This massive BH can then undergo a dynamical capture/exchange 
in a dense stellar environment (e.g., in a globular cluster) placing it in a new, 
massive binary. This binary generates the second BH-BH merger, in which one BH is 
very massive. In this scenario, the capture/exchange rate may be limited by the 
first merger natal kick that could potentially remove the massive BH from a cluster 
environment (Giersz et al. 2015). There is so far no published probability/rate estimate 
for such a specific scenario.

   \end{enumerate}

\begin{acknowledgements}
We would like to thank Sung-Chul Yoon, Mirek Giersz, John Beacom for very useful 
discussions on the subject of this study.  
We would like to thank thousands of {\tt Universe@home} users that have provided their 
personal computers and phones for our simulations, and in particular to Krzysztof 
Piszczek (program IT manager). 
KB acknowledges support from the NCN grant Sonata Bis 2 (DEC-2012/07/E/ST9/01360)
and the NCN grant OPUS (2015/19/B/ST9/01099).
AJR is funded by the Australian Research Council Centre of Excellence
for All-sky Astrophysics (CAASTRO) through Grant
No. CE110001020.
KB is thankful for support from CAASTRO and the Stromlo Distinguished Visitor Program
at the Research School of Astronomy and Astrophysics, Australian National University, 
Canberra. 
AH acknowledges support by the Australian Research Council through an ARC
Future Fellowship FT120100363 and by the US National Science Foundation
under Grant No.~PHY-1430152 (JINA Center for the Evolution of the Elements). 
TB is grateful for support by the NCN grant UMO-2014/15/Z/ST9/00038.
RO acknowledges support from Grants: NSF PHY 1505629 and (via subcontract) AST 1412449.
DEH was supported by NSF CAREER grant PHY-1151836. He also acknowledges support
from the Kavli Institute for Cosmological Physics at the University of Chicago through 
NSF grant PHY-1125897 as well as an endowment from the Kavli Foundation.
EB was supported by NSF CAREER grant PHY-1055103 and by FCT contract
IF/00797/2014/CP1214/CT0012 under the IF2014 Programme. This work was supported by 
the H2020-MSCA-RISE-2015 Grant No. StronGrHEP-690904.
\end{acknowledgements}

\bibliographystyle{aa}
\bibliography{scibib}

\begin{thebibliography}{101}
\expandafter\ifx\csname natexlab\endcsname\relax\def\natexlab#1{#1}\fi

\bibitem[{{Abbott} {et~al.}(2016{\natexlab{a}}){Abbott}, {Abbott}, {Abbott},
  {Abernathy}, {Acernese}, {Ackley}, {Adams}, {Adams}, {Addesso}, {Adhikari},
  \& et~al.}]{AstroPaper}
{Abbott}, B.~P., {Abbott}, R., {Abbott}, T.~D., {et~al.} 2016{\natexlab{a}},
  \apjl, 818, L22

\bibitem[{{Abbott} {et~al.}(2016{\natexlab{b}}){Abbott}, {Abbott}, {Abbott},
  {Abernathy}, {Acernese}, {Ackley}, {Adams}, {Adams}, {Addesso}, {Adhikari},
  \& et~al.}]{LigoO1a}
{Abbott}, B.~P., {Abbott}, R., {Abbott}, T.~D., {et~al.} 2016{\natexlab{b}},
  Physical Review Letters, 116, 241103

\bibitem[{{Bailyn} {et~al.}(1998){Bailyn}, {Jain}, {Coppi}, \&
  {Orosz}}]{Bailyn1998}
{Bailyn}, C.~D., {Jain}, R.~K., {Coppi}, P., \& {Orosz}, J.~A. 1998, \apj, 499,
  367

\bibitem[{{Barkat} {et~al.}(1967){Barkat}, {Rakavy}, \& {Sack}}]{Barkat1967}
{Barkat}, Z., {Rakavy}, G., \& {Sack}, N. 1967, Physical Review Letters, 18,
  379

\bibitem[{{Belczynski} {et~al.}(2010{\natexlab{a}}){Belczynski}, {Bulik},
  {Fryer}, {Ruiter}, {Valsecchi}, {Vink}, \& {Hurley}}]{Belczynski2010b}
{Belczynski}, K., {Bulik}, T., {Fryer}, C.~L., {et~al.} 2010{\natexlab{a}},
  \apj, 714, 1217

\bibitem[{{Belczynski} {et~al.}(2004){Belczynski}, {Bulik}, \&
  {Rudak}}]{Belczynski2004}
{Belczynski}, K., {Bulik}, T., \& {Rudak}, B. 2004, \apjl, 608, L45

\bibitem[{{Belczynski} {et~al.}(2014){Belczynski}, {Buonanno}, {Cantiello},
  {Fryer}, {Holz}, {Mandel}, {Miller}, \& {Walczak}}]{Belczynski2014}
{Belczynski}, K., {Buonanno}, A., {Cantiello}, M., {et~al.} 2014, \apj, 789,
  120

\bibitem[{{Belczynski} {et~al.}(2010{\natexlab{b}}){Belczynski}, {Dominik},
  {Bulik}, {O'Shaughnessy}, {Fryer}, \& {Holz}}]{Belczynski2010a}
{Belczynski}, K., {Dominik}, M., {Bulik}, T., {et~al.} 2010{\natexlab{b}},
  \apjl, 715, L138

\bibitem[{{Belczynski} {et~al.}(2016{\natexlab{a}}){Belczynski}, {Holz},
  {Bulik}, \& {O'Shaughnessy}}]{Belczynski2016b}
{Belczynski}, K., {Holz}, D.~E., {Bulik}, T., \& {O'Shaughnessy}, R.
  2016{\natexlab{a}}, \nat, 534, 512

\bibitem[{{Belczynski} {et~al.}(2002){Belczynski}, {Kalogera}, \&
  {Bulik}}]{Belczynski2002}
{Belczynski}, K., {Kalogera}, V., \& {Bulik}, T. 2002, \apj, 572, 407

\bibitem[{{Belczynski} {et~al.}(2008){Belczynski}, {Kalogera}, {Rasio}, {Taam},
  {Zezas}, {Bulik}, {Maccarone}, \& {Ivanova}}]{Belczynski2008a}
{Belczynski}, K., {Kalogera}, V., {Rasio}, F.~A., {et~al.} 2008, \apjs, 174,
  223

\bibitem[{{Belczynski} {et~al.}(2016{\natexlab{b}}){Belczynski}, {Repetto},
  {Holz}, {O'Shaughnessy}, {Bulik}, {Berti}, {Fryer}, \&
  {Dominik}}]{Belczynski2016a}
{Belczynski}, K., {Repetto}, S., {Holz}, D.~E., {et~al.} 2016{\natexlab{b}},
  \apj, 819, 108

\bibitem[{{Belczynski} \& {Taam}(2008)}]{Belczynski2008c}
{Belczynski}, K. \& {Taam}, R.~E. 2008, \apj, 685, 400

\bibitem[{{Belczynski} {et~al.}(2007){Belczynski}, {Taam}, {Kalogera}, {Rasio},
  \& {Bulik}}]{Belczynski2007}
{Belczynski}, K., {Taam}, R.~E., {Kalogera}, V., {Rasio}, F.~A., \& {Bulik}, T.
  2007, \apj, 662, 504

\bibitem[{{Belczynski} {et~al.}(2012){Belczynski}, {Wiktorowicz}, {Fryer},
  {Holz}, \& {Kalogera}}]{Belczynski2012}
{Belczynski}, K., {Wiktorowicz}, G., {Fryer}, C.~L., {Holz}, D.~E., \&
  {Kalogera}, V. 2012, \apj, 757, 91

\bibitem[{{Bond} {et~al.}(1984){Bond}, {Arnett}, \& {Carr}}]{Bond1984}
{Bond}, J.~R., {Arnett}, W.~D., \& {Carr}, B.~J. 1984, \apj, 280, 825

\bibitem[{{Brown} {et~al.}(2001){Brown}, {Heger}, {Langer}, {Lee}, {Wellstein},
  \& {Bethe}}]{Brown2001}
{Brown}, G.~E., {Heger}, A., {Langer}, N., {et~al.} 2001, \na, 6, 457

\bibitem[{{Chatzopoulos} \& {Wheeler}(2012{\natexlab{a}})}]{Chatzopoulos2012a}
{Chatzopoulos}, E. \& {Wheeler}, J.~C. 2012{\natexlab{a}}, \apj, 748, 42

\bibitem[{{Chatzopoulos} \& {Wheeler}(2012{\natexlab{b}})}]{Chatzopoulos2012b}
{Chatzopoulos}, E. \& {Wheeler}, J.~C. 2012{\natexlab{b}}, \apj, 760, 154

\bibitem[{{Chatzopoulos} {et~al.}(2013){Chatzopoulos}, {Wheeler}, \&
  {Couch}}]{Chatzopoulos2013}
{Chatzopoulos}, E., {Wheeler}, J.~C., \& {Couch}, S.~M. 2013, \apj, 776, 129

\bibitem[{{Chen} {et~al.}(2014){Chen}, {Woosley}, {Heger}, {Almgren}, \&
  {Whalen}}]{Chen2014}
{Chen}, K.-J., {Woosley}, S., {Heger}, A., {Almgren}, A., \& {Whalen}, D.~J.
  2014, \apj, 792, 28

\bibitem[{{Chen} {et~al.}(2015){Chen}, {Bressan}, {Girardi}, {Marigo}, {Kong},
  \& {Lanza}}]{Chen2015}
{Chen}, Y., {Bressan}, A., {Girardi}, L., {et~al.} 2015, \mnras, 452, 1068

\bibitem[{{Choi} {et~al.}(2016){Choi}, {Dotter}, {Conroy}, {Cantiello},
  {Paxton}, \& {Johnson}}]{Choi2016}
{Choi}, J., {Dotter}, A., {Conroy}, C., {et~al.} 2016, \apj, 823, 102

\bibitem[{{Crowther} {et~al.}(2010){Crowther}, {Schnurr}, {Hirschi}, {Yusof},
  {Parker}, {Goodwin}, \& {Kassim}}]{Crowther2010}
{Crowther}, P.~A., {Schnurr}, O., {Hirschi}, R., {et~al.} 2010, \mnras, 408,
  731

\bibitem[{{de Mink} \& {Belczynski}(2015)}]{deMink2015}
{de Mink}, S.~E. \& {Belczynski}, K. 2015, \apj, 814, 58

\bibitem[{{de Mink} \& {Mandel}(2016)}]{deMink2016}
{de Mink}, S.~E. \& {Mandel}, I. 2016, ArXiv e-prints

\bibitem[{{Dessart} {et~al.}(2007){Dessart}, {Burrows}, {Livne}, \&
  {Ott}}]{Dessart2007}
{Dessart}, L., {Burrows}, A., {Livne}, E., \& {Ott}, C.~D. 2007, \apj, 669, 585

\bibitem[{{Dominik} {et~al.}(2012){Dominik}, {Belczynski}, {Fryer}, {Holz},
  {Berti}, {Bulik}, {Mandel}, \& {O'Shaughnessy}}]{Dominik2012}
{Dominik}, M., {Belczynski}, K., {Fryer}, C., {et~al.} 2012, \apj, 759, 52

\bibitem[{{Dominik} {et~al.}(2013){Dominik}, {Belczynski}, {Fryer}, {Holz},
  {Berti}, {Bulik}, {Mandel}, \& {O'Shaughnessy}}]{Dominik2013}
{Dominik}, M., {Belczynski}, K., {Fryer}, C., {et~al.} 2013, \apj, 779, 72

\bibitem[{{Dominik} {et~al.}(2015){Dominik}, {Berti}, {O'Shaughnessy},
  {Mandel}, {Belczynski}, {Fryer}, {Holz}, {Bulik}, \&
  {Pannarale}}]{Dominik2015}
{Dominik}, M., {Berti}, E., {O'Shaughnessy}, R., {et~al.} 2015, \apj, 806, 263

\bibitem[{{Dvorkin} {et~al.}(2015){Dvorkin}, {Silk}, {Vangioni}, {Petitjean},
  \& {Olive}}]{Dvorkin2015}
{Dvorkin}, I., {Silk}, J., {Vangioni}, E., {Petitjean}, P., \& {Olive}, K.~A.
  2015, \mnras, 452, L36

\bibitem[{{Dvorkin} {et~al.}(2016){Dvorkin}, {Vangioni}, {Silk}, {Uzan}, \&
  {Olive}}]{Dvorkin2016}
{Dvorkin}, I., {Vangioni}, E., {Silk}, J., {Uzan}, J.-P., \& {Olive}, K.~A.
  2016, ArXiv e-prints [\eprint[arXiv]{1604.04288}]

\bibitem[{{Eldridge} \& {Stanway}(2016)}]{Eldridge2016}
{Eldridge}, J.~J. \& {Stanway}, E.~R. 2016, ArXiv e-prints

\bibitem[{{Eroshenko}(2016)}]{Eroshenko2016}
{Eroshenko}, Y.~N. 2016, ArXiv e-prints [\eprint[arXiv]{1604.04932}]

\bibitem[{{Ertl} {et~al.}(2016){Ertl}, {Janka}, {Woosley}, {Sukhbold}, \&
  {Ugliano}}]{Ertl2016}
{Ertl}, T., {Janka}, H.-T., {Woosley}, S.~E., {Sukhbold}, T., \& {Ugliano}, M.
  2016, \apj, 818, 124

\bibitem[{{Flanagan} \& {Hughes}(1998)}]{Flanagan1998}
{Flanagan}, {\'E}.~{\'E}. \& {Hughes}, S.~A. 1998, \prd, 57, 4535

\bibitem[{{Fryer}(2014)}]{Fryer2014}
{Fryer}, C. 2014, in Proceedings of Frontier Research in Astrophysics
  (FRAPWS2014) held 26-31 May, 2014 in Mondello (Palermo), Italy. Online at <A
  href=''http://pos.sissa.it/cgi-bin/reader/conf.cgi?confid=237''>http://pos.sissa.it/cgi-bin/reader/conf.cgi?confid=237</A>,
  id.4, 4

\bibitem[{{Fryer}(2006)}]{Fryer2006}
{Fryer}, C.~L. 2006, \nar, 50, 492

\bibitem[{{Fryer} {et~al.}(2012){Fryer}, {Belczynski}, {Wiktorowicz},
  {Dominik}, {Kalogera}, \& {Holz}}]{Fryer2012}
{Fryer}, C.~L., {Belczynski}, K., {Wiktorowicz}, G., {et~al.} 2012, \apj, 749,
  91

\bibitem[{{Fryer} {et~al.}(2001){Fryer}, {Woosley}, \& {Heger}}]{Fryer2001}
{Fryer}, C.~L., {Woosley}, S.~E., \& {Heger}, A. 2001, \apj, 550, 372

\bibitem[{{Georgy} {et~al.}(2013){Georgy}, {Ekstr{\"o}m}, {Eggenberger},
  {Meynet}, {Haemmerl{\'e}}, {Maeder}, {Granada}, {Groh}, {Hirschi}, {Mowlavi},
  {Yusof}, {Charbonnel}, {Decressin}, \& {Barblan}}]{Georgy2013}
{Georgy}, C., {Ekstr{\"o}m}, S., {Eggenberger}, P., {et~al.} 2013, \aap, 558,
  A103

\bibitem[{{Hartwig} {et~al.}(2016){Hartwig}, {Volonteri}, {Bromm}, {Klessen},
  {Barausse}, {Magg}, \& {Stacy}}]{Hartwig2016}
{Hartwig}, T., {Volonteri}, M., {Bromm}, V., {et~al.} 2016, \mnras
  [\eprint[arXiv]{1603.05655}]

\bibitem[{{Heger} {et~al.}(2003){Heger}, {Fryer}, {Woosley}, {Langer}, \&
  {Hartmann}}]{Heger2003}
{Heger}, A., {Fryer}, C.~L., {Woosley}, S.~E., {Langer}, N., \& {Hartmann},
  D.~H. 2003, \apj, 591, 288

\bibitem[{{Heger} \& {Woosley}(2002)}]{Heger2002}
{Heger}, A. \& {Woosley}, S.~E. 2002, \apj, 567, 532

\bibitem[{{Hobbs} {et~al.}(2005){Hobbs}, {Lorimer}, {Lyne}, \&
  {Kramer}}]{Hobbs2005}
{Hobbs}, G., {Lorimer}, D.~R., {Lyne}, A.~G., \& {Kramer}, M. 2005, \mnras,
  360, 974

\bibitem[{{Horiuchi} {et~al.}(2011){Horiuchi}, {Beacom}, {Kochanek}, {Prieto},
  {Stanek}, \& {Thompson}}]{Horiuchi2011}
{Horiuchi}, S., {Beacom}, J.~F., {Kochanek}, C.~S., {et~al.} 2011, \apj, 738,
  154

\bibitem[{{Hujeirat}(2016)}]{Hujeirat2016}
{Hujeirat}, A.~A. 2016, ArXiv e-prints [\eprint[arXiv]{1604.07882}]

\bibitem[{{Hurley} {et~al.}(2000){Hurley}, {Pols}, \& {Tout}}]{Hurley2000}
{Hurley}, J.~R., {Pols}, O.~R., \& {Tout}, C.~A. 2000, \mnras, 315, 543

\bibitem[{{Husa} {et~al.}(2015){Husa}, {Khan}, {Hannam}, {P{\"u}rrer}, {Ohme},
  {Jim{\'e}nez Forteza}, \& {Boh{\'e}}}]{Husa2015}
{Husa}, S., {Khan}, S., {Hannam}, M., {et~al.} 2015, ArXiv e-prints

\bibitem[{{Inayoshi} {et~al.}(2016){Inayoshi}, {Kashiyama}, {Visbal}, \&
  {Haiman}}]{Inayoshi2016}
{Inayoshi}, K., {Kashiyama}, K., {Visbal}, E., \& {Haiman}, Z. 2016, ArXiv
  e-prints [\eprint[arXiv]{1603.06921}]

\bibitem[{{Khan} {et~al.}(2015){Khan}, {Husa}, {Hannam}, {Ohme}, {P{\"u}rrer},
  {Jim{\'e}nez Forteza}, \& {Boh{\'e}}}]{Khan2015}
{Khan}, S., {Husa}, S., {Hannam}, M., {et~al.} 2015, ArXiv e-prints

\bibitem[{{Kinugawa} {et~al.}(2014){Kinugawa}, {Inayoshi}, {Hotokezaka},
  {Nakauchi}, \& {Nakamura}}]{Kinugawa2014}
{Kinugawa}, T., {Inayoshi}, K., {Hotokezaka}, K., {Nakauchi}, D., \&
  {Nakamura}, T. 2014, \mnras, 442, 2963

\bibitem[{{Kistler} {et~al.}(2009){Kistler}, {Y{\"u}ksel}, {Beacom}, {Hopkins},
  \& {Wyithe}}]{Kistler2009}
{Kistler}, M.~D., {Y{\"u}ksel}, H., {Beacom}, J.~F., {Hopkins}, A.~M., \&
  {Wyithe}, J.~S.~B. 2009, \apjl, 705, L104

\bibitem[{{Kobulnicky} {et~al.}(2014){Kobulnicky}, {Kiminki}, {Lundquist},
  {Burke}, {Chapman}, {Keller}, {Lester}, {Rolen}, {Topel}, {Bhattacharjee},
  {Smullen}, {Vargas {\'A}lvarez}, {Runnoe}, {Dale}, \&
  {Brotherton}}]{Kobulnicky2014}
{Kobulnicky}, H.~A., {Kiminki}, D.~C., {Lundquist}, M.~J., {et~al.} 2014,
  \apjs, 213, 34

\bibitem[{{Kreidberg} {et~al.}(2012){Kreidberg}, {Bailyn}, {Farr}, \&
  {Kalogera}}]{Kreidberg2012}
{Kreidberg}, L., {Bailyn}, C.~D., {Farr}, W.~M., \& {Kalogera}, V. 2012, \apj,
  757, 36

\bibitem[{{Lattimer} \& {Yahil}(1989)}]{Lattimer1989}
{Lattimer}, J.~M. \& {Yahil}, A. 1989, \apj, 340, 426

\bibitem[{{Limongi} \& {Chieffi}(2006)}]{Limongi2006}
{Limongi}, M. \& {Chieffi}, A. 2006, \apj, 647, 483

\bibitem[{{Lipunov} {et~al.}(1997){Lipunov}, {Postnov}, \&
  {Prokhorov}}]{Lipunov1997}
{Lipunov}, V.~M., {Postnov}, K.~A., \& {Prokhorov}, M.~E. 1997, Astronomy
  Letters, 23, 492

\bibitem[{{Loeb}(2016)}]{Loeb2016}
{Loeb}, A. 2016, \apjl, 819, L21

\bibitem[{{MacLeod} \& {Ramirez-Ruiz}(2015)}]{Macleod2015}
{MacLeod}, M. \& {Ramirez-Ruiz}, E. 2015, \apj, 803, 41

\bibitem[{{Madau} \& {Dickinson}(2014)}]{Madau2014}
{Madau}, P. \& {Dickinson}, M. 2014, \araa, 52, 415

\bibitem[{{Mandel} \& {de Mink}(2016)}]{Mandel2016}
{Mandel}, I. \& {de Mink}, S.~E. 2016, \mnras, 458, 2634

\bibitem[{{Mapelli}(2016)}]{Mapelli2016}
{Mapelli}, M. 2016, \mnras, 459, 3432

\bibitem[{{Mapelli} {et~al.}(2009){Mapelli}, {Colpi}, \&
  {Zampieri}}]{Mapelli2009}
{Mapelli}, M., {Colpi}, M., \& {Zampieri}, L. 2009, \mnras, 395, L71

\bibitem[{{Marchant} {et~al.}(2016){Marchant}, {Langer}, {Podsiadlowski},
  {Tauris}, \& {Moriya}}]{Marchant2016}
{Marchant}, P., {Langer}, N., {Podsiadlowski}, P., {Tauris}, T., \& {Moriya},
  T. 2016, ArXiv e-prints

\bibitem[{{Mennekens} \& {Vanbeveren}(2014)}]{Mennekens2014}
{Mennekens}, N. \& {Vanbeveren}, D. 2014, \aap, 564, A134

\bibitem[{{Mitchell-Wynne} {et~al.}(2015){Mitchell-Wynne}, {Cooray}, {Gong},
  {Ashby}, {Dolch}, {Ferguson}, {Finkelstein}, {Grogin}, {Kocevski},
  {Koekemoer}, {Primack}, \& {Smidt}}]{Mitchell2015}
{Mitchell-Wynne}, K., {Cooray}, A., {Gong}, Y., {et~al.} 2015, Nature
  Communications, 6, 7945

\bibitem[{{Nelemans} {et~al.}(2001){Nelemans}, {Yungelson}, \& {Portegies
  Zwart}}]{Nelemans2001}
{Nelemans}, G., {Yungelson}, L.~R., \& {Portegies Zwart}, S.~F. 2001, \aap,
  375, 890

\bibitem[{{O'Connor} \& {Ott}(2011)}]{Oconnor2011}
{O'Connor}, E. \& {Ott}, C.~D. 2011, \apj, 730, 70

\bibitem[{{{\"O}zel} {et~al.}(2010){{\"O}zel}, {Psaltis}, {Narayan}, \&
  {McClintock}}]{Ozel2010}
{{\"O}zel}, F., {Psaltis}, D., {Narayan}, R., \& {McClintock}, J.~E. 2010,
  \apj, 725, 1918

\bibitem[{{Pavlovskii} {et~al.}(2016){Pavlovskii}, {Ivanova}, {Belczynski}, \&
  {Van}}]{Pavlovskii2016}
{Pavlovskii}, K., {Ivanova}, N., {Belczynski}, K., \& {Van}, K.~X. 2016, ArXiv
  e-prints [\eprint[arXiv]{1606.04921}]

\bibitem[{{Perna} {et~al.}(2016){Perna}, {Lazzati}, \&
  {Giacomazzo}}]{Perna2016}
{Perna}, R., {Lazzati}, D., \& {Giacomazzo}, B. 2016, \apjl, 821, L18

\bibitem[{{Podsiadlowski} {et~al.}(1992){Podsiadlowski}, {Joss}, \&
  {Hsu}}]{Podsiadlowski1992}
{Podsiadlowski}, P., {Joss}, P.~C., \& {Hsu}, J.~J.~L. 1992, \apj, 391, 246

\bibitem[{{Poelarends} {et~al.}(2008){Poelarends}, {Herwig}, {Langer}, \&
  {Heger}}]{Poelarends2008}
{Poelarends}, A.~J.~T., {Herwig}, F., {Langer}, N., \& {Heger}, A. 2008, \apj,
  675, 614

\bibitem[{{Postnov} \& {Yungelson}(2006)}]{Postnov2006}
{Postnov}, K.~A. \& {Yungelson}, L.~R. 2006, Living Reviews in Relativity, 9
  [\eprint{astro-ph/0701059}]

\bibitem[{{Ricker} \& {Taam}(2008)}]{Ricker2008}
{Ricker}, P.~M. \& {Taam}, R.~E. 2008, \apjl, 672, L41

\bibitem[{{Rodriguez} {et~al.}(2016){Rodriguez}, {Chatterjee}, \&
  {Rasio}}]{Rodriguez2016}
{Rodriguez}, C.~L., {Chatterjee}, S., \& {Rasio}, F.~A. 2016, ArXiv e-prints

\bibitem[{{Rodriguez} {et~al.}(2015){Rodriguez}, {Morscher}, {Pattabiraman},
  {Chatterjee}, {Haster}, \& {Rasio}}]{Rodriguez2015}
{Rodriguez}, C.~L., {Morscher}, M., {Pattabiraman}, B., {et~al.} 2015, Physical
  Review Letters, 115, 051101

\bibitem[{{Sana} {et~al.}(2012){Sana}, {de Mink}, {de Koter}, {Langer},
  {Evans}, {Gieles}, {Gosset}, {Izzard}, {Le Bouquin}, \&
  {Schneider}}]{Sana2012}
{Sana}, H., {de Mink}, S.~E., {de Koter}, A., {et~al.} 2012, Science, 337, 444

\bibitem[{{Sasaki} {et~al.}(2016){Sasaki}, {Suyama}, {Tanaka}, \&
  {Yokoyama}}]{Sasaki2016}
{Sasaki}, M., {Suyama}, T., {Tanaka}, T., \& {Yokoyama}, S. 2016, ArXiv
  e-prints [\eprint[arXiv]{1603.08338}]

\bibitem[{{Schutz}(1989)}]{Schutz1989}
{Schutz}, B.~F. 1989, in NASA Conference Publication, Vol. 3046, NASA
  Conference Publication, ed. R.~W. {Hellings}, 7--13

\bibitem[{{Spera} {et~al.}(2016){Spera}, {Giacobbo}, \& {Mapelli}}]{Spera2016}
{Spera}, M., {Giacobbo}, N., \& {Mapelli}, M. 2016, ArXiv e-prints
  [\eprint[arXiv]{1606.03349}]

\bibitem[{{Spera} {et~al.}(2015){Spera}, {Mapelli}, \& {Bressan}}]{Spera2015}
{Spera}, M., {Mapelli}, M., \& {Bressan}, A. 2015, \mnras, 451, 4086

\bibitem[{{Strolger} {et~al.}(2004){Strolger}, {Riess}, {Dahlen}, {Livio},
  {Panagia}, {Challis}, {Tonry}, {Filippenko}, {Chornock}, {Ferguson},
  {Koekemoer}, {Mobasher}, {Dickinson}, {Giavalisco}, {Casertano}, {Hook},
  {Blondin}, {Leibundgut}, {Nonino}, {Rosati}, {Spinrad}, {Steidel}, {Stern},
  {Garnavich}, {Matheson}, {Grogin}, {Hornschemeier}, {Kretchmer}, {Laidler},
  {Lee}, {Lucas}, {de Mello}, {Moustakas}, {Ravindranath}, {Richardson}, \&
  {Taylor}}]{Strolger2004}
{Strolger}, L.-G., {Riess}, A.~G., {Dahlen}, T., {et~al.} 2004, \apj, 613, 200

\bibitem[{{The LIGO Scientific Collaboration} {et~al.}(2013){The LIGO
  Scientific Collaboration}, {the Virgo Collaboration}, {Abbott}, {Abbott},
  {Abbott}, {Abernathy}, {Acernese}, {Ackley}, {Adams}, {Adams}, \&
  et~al.}]{LVC2013}
{The LIGO Scientific Collaboration}, {the Virgo Collaboration}, {Abbott},
  B.~P., {et~al.} 2013, ArXiv e-prints

\bibitem[{{The LIGO Scientific Collaboration} {et~al.}(2016){The LIGO
  Scientific Collaboration}, {the Virgo Collaboration}, {Abbott}, {Abbott},
  {Abbott}, {Abernathy}, {Acernese}, {Ackley}, {Adams}, {Adams}, \&
  et~al.}]{LigoO1b}
{The LIGO Scientific Collaboration}, {the Virgo Collaboration}, {Abbott},
  B.~P., {et~al.} 2016, ArXiv e-prints [\eprint[arXiv]{1606.04856}]

\bibitem[{{Thorne}(1987)}]{Thorne1987}
{Thorne}, K.~S. 1987, {Gravitational radiation.}, ed. S.~W. {Hawking} \&
  W.~{Israel}, 330--458

\bibitem[{{Timmes} {et~al.}(1996){Timmes}, {Woosley}, \& {Weaver}}]{Timmes1996}
{Timmes}, F.~X., {Woosley}, S.~E., \& {Weaver}, T.~A. 1996, \apj, 457, 834

\bibitem[{{Tutukov} \& {Yungelson}(1993)}]{Tutukov1993}
{Tutukov}, A.~V. \& {Yungelson}, L.~R. 1993, \mnras, 260, 675

\bibitem[{{Vangioni} {et~al.}(2015){Vangioni}, {Olive}, {Prestegard}, {Silk},
  {Petitjean}, \& {Mandic}}]{Vangioni2015}
{Vangioni}, E., {Olive}, K.~A., {Prestegard}, T., {et~al.} 2015, \mnras, 447,
  2575

\bibitem[{{Villante} {et~al.}(2014){Villante}, {Serenelli}, {Delahaye}, \&
  {Pinsonneault}}]{Villante2014}
{Villante}, F.~L., {Serenelli}, A.~M., {Delahaye}, F., \& {Pinsonneault}, M.~H.
  2014, \apj, 787, 13

\bibitem[{{Vink}(2011)}]{Vink2011}
{Vink}, J.~S. 2011, \apss, 336, 163

\bibitem[{{Voss} \& {Tauris}(2003)}]{Voss2003}
{Voss}, R. \& {Tauris}, T.~M. 2003, \mnras, 342, 1169

\bibitem[{{Woosley}(2016)}]{Woosley2016}
{Woosley}, S.~E. 2016, \apjl, 824, L10

\bibitem[{{Woosley} {et~al.}(2007){Woosley}, {Blinnikov}, \&
  {Heger}}]{Woosley2007}
{Woosley}, S.~E., {Blinnikov}, S., \& {Heger}, A. 2007, \nat, 450, 390

\bibitem[{{Woosley} \& {Heger}(2015)}]{Woosley2015}
{Woosley}, S.~E. \& {Heger}, A. 2015, in Astrophysics and Space Science
  Library, Vol. 412, Very Massive Stars in the Local Universe, ed. J.~S.
  {Vink}, 199

\bibitem[{{Woosley} {et~al.}(2002){Woosley}, {Heger}, \&
  {Weaver}}]{Woosley2002}
{Woosley}, S.~E., {Heger}, A., \& {Weaver}, T.~A. 2002, Reviews of Modern
  Physics, 74, 1015

\bibitem[{{Xu} \& {Li}(2010)}]{Xu2010}
{Xu}, X.-J. \& {Li}, X.-D. 2010, \apj, 722, 1985

\bibitem[{{Young} {et~al.}(2009){Young}, {Ellinger}, {Arnett}, {Fryer}, \&
  {Rockefeller}}]{Young2009}
{Young}, P.~A., {Ellinger}, C.~I., {Arnett}, D., {Fryer}, C.~L., \&
  {Rockefeller}, G. 2009, \apj, 699, 938

\bibitem[{{Yusof} {et~al.}(2013){Yusof}, {Hirschi}, {Meynet}, {Crowther},
  {Ekstr{\"o}m}, {Frischknecht}, {Georgy}, {Abu Kassim}, \&
  {Schnurr}}]{Yusof2013}
{Yusof}, N., {Hirschi}, R., {Meynet}, G., {et~al.} 2013, \mnras, 433, 1114

\bibitem[{{Zampieri} \& {Roberts}(2009)}]{Zampieri2009}
{Zampieri}, L. \& {Roberts}, T.~P. 2009, \mnras, 400, 677

\end{thebibliography}

\end{document}